\documentclass[preprint,showpacs,preprintnumbers,amsmath,amssymb,nofootinbib]{revtex4}
\usepackage{booktabs}
\usepackage{mathrsfs}
\usepackage{epsfig}
\usepackage{graphicx}
\usepackage{dcolumn}
\usepackage{bm}
\usepackage{amsmath}
\usepackage{slashed}
\usepackage{multirow}

\let\jnfont=\rm
\def\NPB#1,{{\jnfont Nucl.\ Phys.\ B }{\bf #1},}
\def\PLB#1,{{\jnfont Phys.\ Lett.\ B }{\bf #1},}
\def\EPJC#1,{{\jnfont Eur.\ Phys.\ Jour.\ C }{\bf #1},}
\def\PRD#1,{{\jnfont Phys.\ Rev.\ D }{\bf #1},}
\def\PRL#1,{{\jnfont Phys.\ Rev.\ Lett.\ }{\bf #1},}
\def\MPLA#1,{{\jnfont Mod.\ Phys.\ Lett.\ A }{\bf #1},}
\def\JPG#1,{{\jnfont J.\ Phys.\ G}{\bf #1},}
\def\CTP#1,{{\jnfont Commun.\ Theor.\ Phys.\ }{\bf #1},}
\def\ZPC#1,{{\jnfont Z.\ Phys.\ C }{\bf #1},}
\def\JHEP#1,{{\jnfont JHEP \ }{\bf #1},}
\def\Rv{\not{\hbox{\kern-1pt $R$}}}
\def\p{\not{\hbox{\kern-3pt $p$}}}

\newcommand{\bea}{\begin{eqnarray}}
\newcommand{\eea}{\end{eqnarray}}

\newcommand{\bcen}{\begin{center}}
\newcommand{\ecen}{\end{center}}

\newcommand{\beq}{\begin{eqnarray}}
\newcommand{\eeq}{\end{eqnarray}}

\def\t1{\tilde{t_1}}

\begin{document}

\title{Closing up a light stop window in natural SUSY at LHC}
\author{ Archil Kobakhidze$^{1}$}
\author{ Ning Liu$^{2}$}
\author{ Lei Wu$^{2}$}
\author{ Jin Min Yang$^{3,4}$}
\author{ Mengchao Zhang$^{4}$}
\affiliation{$^1$ ARC Centre of Excellence for Particle Physics at the Terascale, School of Physics, The University of Sydney, NSW 2006, Australia\\
$^2$ Institution of Theoretical Physics, Henan Normal University, Xinxiang 453007, China\\
$^3$ Department of Physics, Tohoku University, Sendai 980-8578, Japan\\
$^4$ Institute of Theoretical Physics, Academia Sinica, Beijing 100190, China
   \vspace*{1.5cm} }%

\date{\today}

\begin{abstract}

Top squark (stop) plays a key role in the radiative stability of the Higgs boson mass in supersymmetry
 (SUSY). In this work, we use the LHC Run-1 data to determine the lower mass limit of the right-handed
stop in a natural SUSY scenario, where the higgsinos $\tilde{\chi}^0_{1,2}$ and $\tilde{\chi}^\pm_{1}$ are
light and nearly degenerate. We find that the stop mass has been excluded up
to 430 GeV for $m_{\tilde{\chi}^0_1} \lesssim 250$ GeV and to 540 GeV for $m_{\tilde{\chi}^0_1} \simeq 100$ GeV by the Run-1 SUSY searches for $2b+E^{miss}_T$ and $1\ell+jets+E^{miss}_T$, respectively. In a small strip of parameter space with $m_{\tilde{\chi}^0_1} \gtrsim 190$ GeV, the stop mass can still be as light as 210 GeV and compatible with the Higgs mass measurement and the monojet bound. The 14 TeV LHC with a luminosity of 20 fb$^{-1}$ can further cover such a light stop window by monojet and $2b+E^{miss}_T$ searches and push the lower bound of the stop mass to 710 GeV.
We also explore the potential to use the Higgs golden ratio,
$D_{\gamma\gamma}=\sigma(pp \to h \to \gamma\gamma)/\sigma(pp \to h \to ZZ^* \to4\ell^\pm)$, as a complementary
probe for the light and compressed stop. If this golden ratio can be measured at percent
level at the high luminosity LHC (HL-LHC) or future $e^+e^-$ colliders, the light stop can be excluded
for most of the currently allowed parameter region.

\end{abstract}

\maketitle

\section{INTRODUCTION}

Weak scale supersymmetry is a leading candidate for solving the naturalness problem of the Standard Model, i.e.
explaining the radiative stability of the hierarchy between the electroweak scale and high energy scales,
such as Planck mass. In the minimal supersymmetric standard model (MSSM), the minimization
condition of the Higgs potential is given by \cite{mz}
\begin{eqnarray}
\frac{M^2_{Z}}{2}&=&\frac{(m^2_{H_d}+\Sigma_{d})-(m^2_{H_u}+
\Sigma_{u})\tan^{2}\beta}{\tan^{2}\beta-1}-\mu^{2},
\label{minimization}
\end{eqnarray}
where $m^2_{H_d}$ and $m^2_{H_u}$ denote the weak scale soft SUSY breaking masses of the Higgs fields, $\tan\beta = v_u/v_d$ and $\mu$
is the higgsino mass parameter. $\Sigma_{u}$ and $\Sigma_{d}$ arise from the radiative corrections to the tree level Higgs potential, which include the contributions from various particles and sparticles with sizeable Yukawa and/or gauge couplings to the Higgs sector. Explicit forms for the $\Sigma_u$ and $\Sigma_d$ are given in the Appendix of Ref. Obviously, in order to get the observed value of $M_Z$ without finely tuned cancellations in Eq.~(\ref{minimization}), each term on the right-hand side should be comparable in magnitude. This then suggests that the electroweak fine-tuning of $M^2_Z$ can be quantified by $\Delta^{-1}_{EW}$\footnote{The Barbieri and Guidice (BG) measure in Ref.~\cite{bg} is applicable to a theory with several independent effective theory parameters. But for a more fundamental theory, BG measure often leads to over-estimates of fine tuning \cite{baer-ew}.},
\begin{eqnarray}
\Delta^{-1}_{EW}\equiv  (M^2_Z/2)/max_i|C_i|.
\label{ewnaturalness}
\end{eqnarray}
Here, $C_{H_u}=-m^2_{H_u}\tan^2\beta/(\tan^2\beta-1)$, $C_{H_d}=m^2_{H_d}/(\tan^2\beta-1)$ and $C_\mu=-\mu^2$. Also, $C_{\Sigma_u(i)}=-\Sigma_u(i)(\tan^2\beta)/(\tan^\beta-1)$ and $C_{\Sigma_d(i)}=\Sigma_d(i)/(\tan^\beta-1)$, where $i$ labels the various loop
contributions to $\Sigma_u$ and $\Sigma_d$. So an upper bound on $\Delta^{-1}_{EW} \gtrsim 10\%$ from naturalness considerations implies that the higgsino mass parameter $\mu$ must be of the order of $\sim 100-200$ GeV. Hence, to probe the SUSY naturalness at LHC, the most essential task is to search for light higgsinos. However, due to the low (percent level) signal-to-background ratio, detecting the pair production of these nearly degenerate higgsinos through monojet(-like) or vector boson fusion events seems challenging at LHC \cite{higgsino1,higgsino2,higgsino3}.

Besides higgsinos, the stops usually strongly relate with the naturalness, which can contribute to $\Delta^{-1}_{EW}$ at one-loop level and favor the stop mass not to be too heavy\footnote{In some supersymmetric models, such as Ref.~\cite{non-universal-gaugino}, the bound on the stop mass from naturalness can be weakened due to the cancellation between stop loop and other sparticle loops.} \cite{roman}. In addition, there are other good theoretical motivations of considering a light stop. For example, in some popular grand unification models, supersymmetry breaking is usually assumed to transmit to the visible sector at a certain high energy scale, and then Yukawa contributions to the renormalization group evolution tend to reduce stop masses more than other squark masses. Another one is that the chiral mixing for certain flavor squarks is proportional to the mass of the corresponding quark, and is therefore more sizable for stops. Such a mixing will further reduce the mass of the lighter stop. Moreover, we note that a light stop is phenomenologically needed by the electroweak baryogenesis \cite{baryogenesis}. Given these, the searches for pair/single production of stop are also important to understand the naturalness and to test supersymmetric models at LHC \cite{rs,baer,wu-1}.

So far, experimental searches for stops at LHC Run-1  have resulted in bounds on stop masses of a few hundred GeV
\cite{atlas-stop-1,atlas-stop-2,atlas-stop-3,atlas-stop-4,atlas-stop-5,atlas-stop-6,cms-stop-1,cms-stop-2,cms-stop-3,cms-stop-4,cms-stop-5}. The present search strategies of the direct stop pair production mainly depend on the mass splitting
between the stop and the lighest supersymmetric partner  (LSP). For example, when $\Delta m_{\tilde{t}_1-\tilde{\chi}^0_1} \gg m_t$,
the top quark from stop decay can be quit energetic as compared with the top quarks in the $t\bar{t}$ background. Therefore,
certain endpoint observables, like $M_{T_2}$, can be used to efficiently reduce the $t\bar{t}$ background
\cite{atlas-stop-2,atlas-stop-4,atlas-stop-5,cms-stop-4,cms-stop-5}. Contrary to this, in the compressed region,
where $\Delta m_{\tilde{t}_1-\tilde{\chi}^0_1}\approx m_t$, the kinematics of the top quarks from stop decay are similar to those
in the top pair production and the standard search strategies often suffer from a poor sensitivity. For this case, one way
is to compare the top pair production cross section measurement with the theoretical prediction, which can rule out stop
masses below $\sim 180$ GeV for a light neutralino LSP \cite{atlas-stop-3,tt-1,tt-2}. Another way is to use a high momentum
jet recoiling against $\tilde{t}_1\tilde{t}^*_1$ system to produce the large $E^{miss}_T$ and anti-correlation between $E^{miss}_T$
and the recoil jet transverse vectors \cite{rest-top-1,rest-top-2,rest-top-3}. Furthermore,
if $\Delta m_{\tilde{t}_1-\tilde{\chi}^0_1} \ll m_t$, the stop decay will be dominated by the four-body channel
$\tilde{t}_1 \to bf'\bar{f}\tilde{\chi}^0_1$ or the two-body loop channel $\tilde{t}_1 \to c \tilde{\chi}^0_1$
\cite{hikasa,djouadi,margrate,andreas}. But due to the small mass difference, the decay products of the stop are usually
too soft to be observed. Thus the single high $p_T$ hard jet from the ISR/FSR (with the heavy quark tagging) is used to
tag these compressed stop events \cite{drees,choudhury,qishu,litong}. At the same time, many theoretical studies have been devoted
to improving the LHC sensitivity to the stop searches in some special kinematical regions \cite{stop-th-1} and to
constraining the light stops in various theoretical frameworks \cite{stop-th-2}.

Besides the sparticle mass splitting, the assumption on the branching ratios of stop and the nature of neutralinos can
significantly affect the sensitivity of the LHC direct searches. For examples, if $M_{1,2}\gg \mu$, the left-handed stop
decay  $\tilde{t}_1 \to t \tilde{\chi}^{0}_{1,2}$ is  enhanced by the large top quark Yukawa coupling. Also, due to the
$SU(2)$ symmetry and nearly degenerate higgsinos ($\tilde{\chi}^0_{1,2}$ and $\tilde{\chi}^\pm_{1}$), the left-handed sbottom
decays $\tilde{b}_1 \to t \tilde{\chi}^-_1$ inevitably mimics the stop signals $\tilde{t}_{1} \to t \tilde{\chi}^{0}_{1,2}$.
The combined null results of the stop and sbottom searches have excluded a left-handed stop below about 600 GeV in natural
SUSY scenario \cite{ruderman,martin,wu-2,warsaw}. On the other hand, since the right-handed stop has no $SU(2)$ gauge symmetry link
with the sbottom sector, sbottoms can be decoupled and will not necessarily contribute to the stop events. Thus, the LHC
direct search constraints on the right-handed stop will become weaker, and may still allow stop mass around the weak scale.

In this work, we use the LHC Run-1 data to determine the lower mass limit of the right-handed stop in a natural SUSY scenario,
where the higgsinos $\tilde{\chi}^0_{1,2}$ and $\tilde{\chi}^\pm_{1}$ are light and nearly degenerate in
mass ( $2{~\rm GeV} \lesssim \Delta m \lesssim 5{~\rm GeV}$). Then we investigate the prospect of closing up
the currently allowed light right-handed stop mass region through the direct searches for
$2b+E^{miss}_{T}$, $1\ell+jets+E^{miss}_T$ and monojet events at 14 TeV LHC. Apart from the direct searches, one may also utilise indirect observations to constrain the light stops. Namely, the light stops can significantly affect the loop processes $gg \to h$ and $h \to \gamma\gamma$. With the upgrade of LHC, the Higgs couplings with the gauge bosons will be measured with much higher experimental accuracy than the current measurements and may be used to indirectly constrain our scenario. We also explore the potential of the Higgs golden ratio
$D_{\gamma\gamma}=\sigma(pp \to h \to \gamma\gamma)/\sigma(pp \to h \to ZZ^* \to4\ell^\pm)$ \cite{higgs-ratio}
as a complementary probe for the light stop scenario.

\section{calculations, results and discussion}\label{section2}

Considering the higgsinos and stops are closely related to the naturalness problem, we scan the following region
of the MSSM parameter space:
\begin{eqnarray}
100~{\rm GeV} \le \mu \le 300~{\rm GeV}, \quad 100~{\rm GeV} \le ~m_{\tilde{t}_R}\le 1~{\rm TeV}, \nonumber \\
1.5~{\rm TeV} \le ~m_{\tilde{Q}_{3L}}\le 3~{\rm TeV}, \quad 1~{\rm TeV} \le A_t \le 3~{\rm TeV}, \quad 5 \le \tan\beta \le 50.
\end{eqnarray}
As our study is performed in a simplified phenomenological MSSM, we abandon the relation $M_1:M_2:M_3=1:2:7$ inspired the gaugino mass unification \footnote{Note that one possible way to relax the naturalness problem is to choose a suitable boundary condition of gaugino masses at the GUT scale, such as $M_2:M_3 \simeq 5:1$ in Ref.~\cite{omura}.} and assume $M_1=M_2=2$ TeV at the weak scale for simplicity. Such a condition leads to the nearly degenerate higgsinos (with the mass splitting around 2-5 GeV). Besides, in order to avoid introducing too much fine-tuning, we take $M_3=1.5$ TeV, which usually contributes to the Higgs mass at two-loop level. The sleptons and the first two generations of squarks in natural SUSY are supposed to be heavy to avoid the SUSY flavor and CP problems, which are all fixed at 3 TeV. We also assume $m_A=1$ TeV, $A_{b}=0$ and $m_{\tilde{b}_R}=2$ TeV. Such a setup will make our lighter stop $\tilde{t}_1$ dominated by the right-handed component, and also provide the correct Higgs mass. In our scan we consider the following constraints:
\subsection{Indirect Constraints}
\begin{itemize}
\item[(1)] We choose the light CP-even Higgs boson as the SM-like Higgs boson and require its mass in the range of 123--127 GeV.
We use the package of \textsf{FeynHiggs-2.11.2} \cite{feynhiggs} to calculate the Higgs mass \footnote{In general, different packages may give a different Higgs mass prediction. It is known from the MSSM that spectrum generators performing a $\bar{DR}$ calculation (such as Suspect \cite{suspect}) can agree quite well, while sizable differences to the OS calculation of FeynHiggs exists. The differences are assumed to arise from the missing electroweak corrections and momentum dependence at two-loop level as well as from the dominant three-loop corrections. These are the effects that underlie the often-quoted estimate of a few GeV uncertainty for the SM-like Higgs mass in the MSSM \cite{higgsmass}.}. Besides, a light stop with the large mixing trilinear parameter $A_t$ needed by the Higgs mass often leads to a global vacuum where charge and colour are broken \cite{color-breaking-1,color-breaking-2}. We impose the constraint of the metastability of the vacuum state by requiring $|A_t| \lesssim 2.67\sqrt{M^2_{\tilde{Q}_{3L}}+M^2_{\tilde{t}_R}+M^2_A \cos^2\beta}$ \cite{color-breaking-2}.
\item[(2)] Since the light stop and higgsinos can contribute to the B-physics observables, we require our samples to satisfy the
bound of $B\rightarrow X_s\gamma$ at 2$\sigma$ level. We use the package of \textsf{SuperIso v3.3} \cite{superiso} to implement
this constraint.
\item[(3)] As known, in the natural MSSM, the thermal relic density of the light higgsino-like neutralino dark matter is
typically low because of the large annihilation rate in the early universe. In order to provide the required relic density,
several alternative ways have been proposed \cite{non-sm-dm1,non-sm-dm2,non-sm-dm3}, such as choosing the axion-higgsino
admixture as the dark matter \cite{axion}. However, if the naturalness requirement is relaxed, the heavy higgsino-like
neutralino with a mass about 1 TeV can solely produce the correct relic density in the MSSM \cite{roy}. So we require the
thermal relic density of the neutralino dark matter is below the 2$\sigma$ upper limit of the Planck value \cite{planck}.
We use the package of \textsf{MicrOmega v2.4} \cite{micromega} to calculate the relic density.
\end{itemize}
We have also verified using  \textsf{HiggsBounds-4.2.1} \cite{higgsbounds} and \textsf{HiggsSignals-1.4.0} \cite{higgssignals}
packages that  the samples allowed by the above constraints are also consistent with the Higgs data from LEP, Tevatron and LHC.

\begin{figure}[ht]
\centering
\includegraphics[width=3.5in]{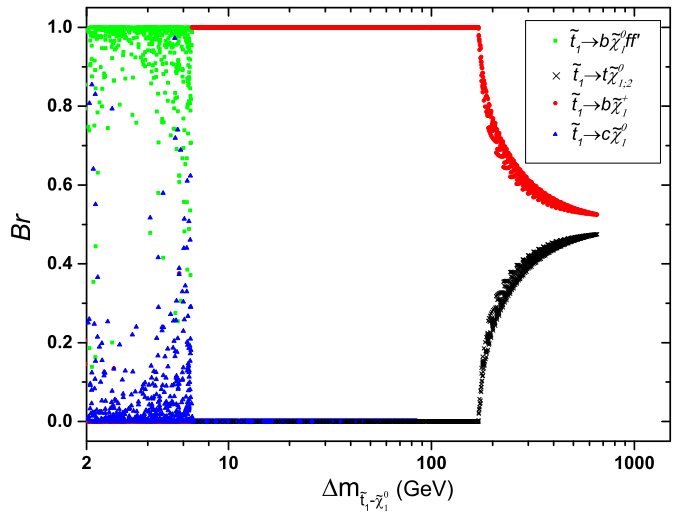}
\vspace{-0.5cm}
\caption{Dependence of the stop decay branching ratio on the mass splitting $\Delta m_{\tilde{t}_1-\tilde{\chi}^0_{1}}$.}
\label{br}
\end{figure}

In Fig.\ref{br}, we show the dependence of the stop decay branching ratios on the mass splitting
$\Delta m_{\tilde{t}_1-\tilde{\chi}^0_{1}}$ in our scenario. The branching ratios are calculated by the package
of \textsf{SDECAY} \cite{sdecay}. We can see that a heavy right-handed stop decays to $b\tilde{\chi}^+_1$
with $Br\simeq 50\%$ and $t\tilde{\chi}^0_{1,2}$ with $Br\simeq 25\%, 25\%$. This is because the partial
decay width $\Gamma (\tilde{t}_1 \to b \tilde{\chi}^+_1)$ and  $\Gamma (\tilde{t}_1 \to t \tilde{\chi}^0_{1,2})$
are both proportional to $y^2_t$ ($y_t$ is the top quark Yukawa coupling) \cite{djouadi}.
Other decay modes $\tilde{t}_1 \to t \tilde{\chi}^0_{3,4}$ are
kinematically forbidden because the bino and wino mass is assumed to be decoupled in our calculations.
For $m_b+m_W< \Delta m_{\tilde{t}_1-\tilde{\chi}^0_{1}} < m_t$, the dominant decay process is still
$\tilde{t}_1 \to b \tilde{\chi}^+_1$ because it has a much larger phase space than the three-body decay channel
$\tilde{t}_1 \to bW\tilde{\chi}^0_{1}$. Further, if $\Delta m_{\tilde{t}_1-\tilde{\chi}^0_{1}} < m_b+m_W$, the four-body
decay process $\tilde{t}_1 \to bf\bar{f}'\tilde{\chi}^0_{1}$ and the loop decay channel $\tilde{t}_1 \to c\tilde{\chi}^0_{1}$
are extremely suppressed (except for the region where $\tilde{t}_1 \to b\tilde{\chi}^+_1$ is kinematically forbidden),
as shown in Fig.\ref{br}. The reason is that our stop is predominantly right-handed and the neutralino $\tilde{\chi}^0_{1}$
is higgsino-like, so that the decay width of $\tilde{t}_1 \to c\tilde{\chi}^0_{1}$ is heavily reduced because of tiny
$\tilde{t}_{L,R}-\tilde{c}_L$ mixing and of the gaugino-higgsino nature of neutralinos \cite{hikasa}. The decay
$\tilde{t}_1 \to bf\bar{f}'\tilde{\chi}^0_{1}$ is also suppressed due to the small phase space (note that the
neutralinos $\tilde{\chi}^0_{1,2}$ and the chargino $\tilde{\chi}^+_1$ are nearly degenerate higgsinos).

\subsection{Direct Constraints}
In our scenario, due to $M_{1,2}\gg \mu$, the higgsinos $\tilde{\chi}^{\pm}_{1}$ and $\tilde{\chi}^0_{1,2}$ are nearly
degenerate so that their decay products are too soft to be tagged at LHC. Such a feature can
change the conventional LHC signatures in some certain stop decay channels. For example, the stop pair production
followed by the dominant decay
$\tilde{t}_1 \to b \tilde{\chi}^+_1$ will appear as $2b+E^{miss}_T$. So in our study, we consider the following
relevant LHC direct search constraints at $\sqrt{s}=8$ TeV:
\begin{itemize}
\item[(1)] The ATLAS search for stop/sbottom pair production in final states with missing transverse momentum and two b-jets \cite{atlas-stop-6}.
\item[(2)] The ATLAS and CMS search for stop pair production in final states with one isolated lepton, jets, and missing
transverse momentum \cite{atlas-stop-2,cms-stop-4};
\item[(3)] The ATLAS search for pair-produced stops decaying to charm quark or in compressed supersymmetric
scenarios \cite{atlas-stop-1}.
\end{itemize}
\begin{table}[t]
\label{tab1}
\caption{The signals of the LHC stop direct searches and the corresponding sources in the natural SUSY.}
\begin{tabular}{|c|c|}
\hline~~ LHC stop direct searches~~&
~~Sources in natural SUSY~~\\
\hline
$\ell+jets+ E^{miss}_{T}$ \cite{atlas-stop-2,cms-stop-4} &
$
\begin{array}{ll} pp\to \tilde{t}_1 \tilde{t}_1~ (\tilde{t}_1 \to t \tilde{\chi}^{0}_{1,2})
   \end{array}
$
 \\ \hline
$2b+E^{miss}_{T}$ \cite{atlas-stop-6}&
$
\begin{array}{ll}
  pp\to \tilde{t}_1 \tilde{t}_1~ (\tilde{t}_1 \to b \tilde{\chi}^{+}_{1})
   \end{array}
$ \\
\hline
$jet+E^{miss}_{T}$ \cite{atlas-stop-1}&
$
\begin{array}{ll}
   pp\to jet+\tilde{t}_1 \tilde{t}_1~ (\tilde{t}_1 \to b \tilde{\chi}^{+}_{1}, b f\bar{f'}\tilde{\chi}^{0}_{1,2}, c\tilde{\chi}^0_{1,2}) \end{array}
$
 \\ \hline
\end{tabular}
\end{table}
In Table \ref{tab1}, we summarize the signals of the above direct searches and the corresponding source of each signal
in our scenario. We use the packages \textsf{CheckMATE-1.2.1} \cite{checkmate-1} and \textsf{MadAnalysis 5-1.1.12} \cite{ma5}
to recast the above ATLAS analyses (1)-(3) and CMS analysis (2), respectively. We calculate the NLO+NLL cross section of
the stop pair production by using \textsf{NLL-fast} package \cite{nll-fast} with the CTEQ6.6M PDFs \cite{cteq6}. The parton level signal events are generated by the package \textsf{MadGraph5} \cite{mad5} and are showered and hadronized by the package \textsf{PYTHIA} \cite{pythia}.
The detector simulation effects are implemented with the tuned package \textsf{Delphes} \cite{delphes}, which is included
in \textsf{CheckMATE-1.2.1} and \textsf{MadAnalysis 5-1.1.12}. The jets are clustered with the anti-$k_t$ algorithm \cite{antikt}
by the package \textsf{FastJet} \cite{fastjet}. Finally, we define the ratio $r = max(N_{S,i}/S^{95\%}_{obs,i})$ for each
experimental search. Here $N_{S,i}$ is the number of the signal events for the $i$-th signal region and $S^{95\%}_{obs,i}$ is
the corresponding observed 95\% C.L. upper limit. The $max$ is over all the signal regions for each search. If $r > 1$,
we conclude that such a point is excluded at 95\% C.L..

\subsection{Results}

\begin{figure}[ht]
\centering
\includegraphics[width=5in]{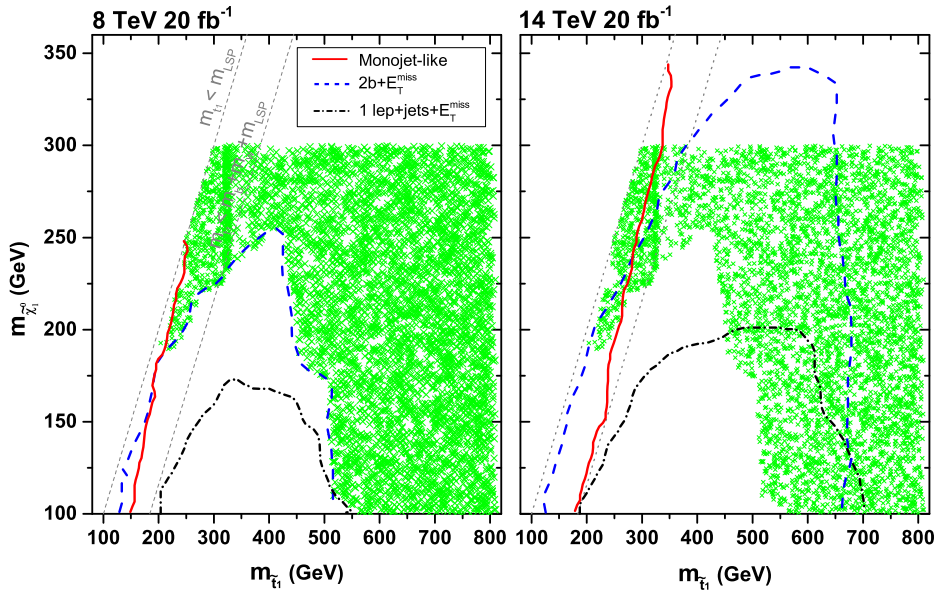}
\vspace*{-.5cm}
\caption{Regions excluded by the direct searches for the stop pair at
8 TeV run (left panel) and extrapolation to the 14 TeV run (right panel) with ${\cal L}=20$ fb$^{-1}$.
For $1\ell+jets+E^{miss}_T$ and $2b+E^{miss}_T$, the region below each curve is the excluded region.
For the monojet search, the region to the left of the curve is its excluded region. The green crosses represent the samples allowed by the current indirect and direct constraints.}
\label{constraints}
\end{figure}
In Fig.\ref{constraints}, we plot the exclusion limits of the direct searches for the stop pair in the plane of $m_{\tilde{t}_1}$
versus LSP mass at 8 TeV LHC with ${\cal L}=20$ fb$^{-1}$. The green crosses represent the samples allowed by the current
indirect and direct constraints. Since the moderate or heavy right-handed stop dominantly decays to $b \tilde{\chi}^+_1$ and
$t\tilde{\chi}^0_{1,2}$, which produces $2b+E^{miss}_T$ and $t\bar{t}+E^{miss}_T$ signatures respectively, we can see that the
searches for $2b+E^{miss}_T$ and $1\ell+jets+E^{miss}_T$ events give strong bounds on the stop mass in the region with
$\Delta m_{\tilde{t}_1-\tilde{\chi}^0_{1}} > m_t$. For example, when $\mu \simeq 100~(250)$ GeV, the stop mass has been excluded up
to about 540 (430) GeV by $1\ell+jets+E^{miss}_T$ ($2b+E^{miss}_T$). If the stop mass is close to the LSP mass, the $b$-jets
from the stop decay $\tilde{t}_1 \to b \tilde{\chi}^+_1/b f\bar{f'}\tilde{\chi}^0_{1,2}$ or $c$-jets from
$\tilde{t}_1 \to c \tilde{\chi}^0_{1,2}$ become soft. Then the monojet search will be a sensitive probe for this region and
can exclude the stop mass up to 150 GeV for $\mu \simeq 100$ GeV. While in a small strip of parameter space with $m_{\tilde{\chi}^0_1} \gtrsim 190$ GeV, the stop mass can still be as light as 210 GeV and compatible with all the current bounds. As pointed in \cite{monojet}, the higher energy will improve the sensitivity of the monojet searches for a mass splitting below 100 GeV. So, we regenerate the corresponding signals and backgrounds, and extrapolate our analyses to 14 TeV LHC by taking the same cut values and the definitions of the signal regions as those at 8 TeV LHC \footnote{Here we conservatively estimate the exclusion limits at 14 TeV LHC. The optimization of the cut values and the signal regions may further improve our results.}. Then, we can see that such a narrow region for a light stop can be further covered by the constraints of monojet and $2b+E^{miss}_T$ with 20 fb$^{-1}$ of data. At the same time, the lower bound of the stop mass will be pushed up to about 660 GeV for
$m_{\tilde{\chi}^0_1} \lesssim 330$ GeV and 710 GeV for $m_{\tilde{\chi}^0_1} \sim 100$ GeV by $2b+E^{miss}_T$ and $1\ell+jets+E^{miss}_T$ searches, respectively.

\begin{figure}[ht]
\centering
\includegraphics[width=3.5in]{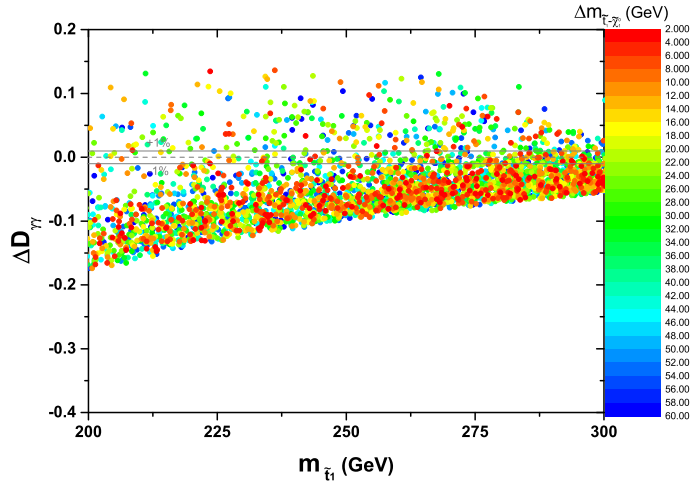}
\vspace*{-.5cm}
\caption{Constraints of the Higgs golden ratio
$D_{\gamma\gamma}=\sigma(pp \to h \to \gamma\gamma)/\sigma(pp \to h \to ZZ^* \to4\ell^\pm)$ on the light stop shown
in the left panel of Fig.\ref{constraints}. The colormap represents the mass difference between the stop and the LSP.}
\label{ratio}
\end{figure}
On the other hand, with more data collected at the LHC, the precision measurement of Higgs couplings can be used as
indirect probes of light new particles. In natural SUSY, the stops may significantly change the loop processes
$gg \to h$ and $h \to \gamma\gamma$. However, the signal strength measurement of $pp \to h \to \gamma\gamma$
suffers from some theoretical uncertainties \cite{higgs-uncertainty}. To solve this problem, a high-precision
Higgs observable $D_{\gamma\gamma}$ that can be measured at percent level was constructed by using the ratio of the
Higgs golden channel signal strengthes \cite{higgs-ratio},
\begin{eqnarray}
D_{\gamma\gamma}=\mu(pp \to h \to \gamma\gamma)/\mu(pp \to h \to ZZ^* \to4\ell^\pm).
\label{drr}
\end{eqnarray}
In Fig.\ref{ratio}, we present the constraints of the Higgs golden ratio $D_{\gamma\gamma}$ on the light stop window shown
in Fig.\ref{constraints}. It can be seen that the stop with mass $m_{\tilde{t}_1} \simeq m_t$ can significantly reduce
the value of $D_{\gamma\gamma}$ by about 18\% because such a light stop will cancel with the contribution of $W$-loop
in the decay of $h \to \gamma\gamma$. While with the increase of the stop mass, the contribution of the stop loop
can change the sign and constructively interfere with the $W$-loop. On the other hand, since the decay width
of $h \to \gamma\gamma$ also depends on the trilinear parameter $A_t$ and $\tan\beta$ \cite{yang}, some of our samples
can make $D_{\gamma\gamma}$ very close to 1. Therefore, if the golden ratio $D_{\gamma\gamma}$ can be measured at 1\% level
(as discussed in \cite{higgs-ratio}) at the HL-LHC or future $e^+e^-$ colliders, most of our light stop region
allowed by 8 TeV LHC can be excluded.

\section{conclusions}
In this work we used the LHC Run-1 data to constrain the right-handed stop in a natural SUSY scenario,
where the higgsinos $\tilde{\chi}^0_{1,2}$ and $\tilde{\chi}^\pm_{1}$ are light ($\mu \simeq 100-300$ GeV)
and nearly degenerate. For $m_{\tilde{t}_1} \gg m_t$, we found that the stop mass is excluded up to about 540 (430) GeV
for $\mu\simeq100~(250)$ GeV by the 8 TeV LHC direct searches in $1\ell+jets+E^{miss}_T$ ($2b+E^{miss}_T$) channel.
However, in a small strip of parameter space with $m_{\tilde{\chi}^0_1} \gtrsim 190$ GeV, the stop mass can still be as light as 210 GeV and compatible with the bounds from the Higgs mass and the current monojet searches. We have extrapolated our analyses to 14 TeV LHC and found that such a light stop mass window can be further covered by the monojet and $2b+E^{miss}_T$ searches. The lower bound of the stop mass will be pushed up to about 710 GeV. We also found that the precision measurement of the Higgs golden ratio $D_{\gamma\gamma}=\sigma(pp \to h \to \gamma\gamma)/\sigma(pp \to h \to ZZ^* \to4\ell^\pm)$ at percent level can exclude most of our light stop region and thus play a complementary role in probing the light stop.

\acknowledgments
We thank Manuel Drees and Jong Song Kim for helpful discussions.
This work is partly supported by the Australian Research Council, by the National Natural Science Foundation of
China (NNSFC) under grants Nos. 11275057, 11305049, 11375001, 11405047, 11135003, 11275245,
by Specialised Research Fund for the Doctoral Program of Higher Education under Grant No. 20134104120002, by the Startup Foundation for Doctors of
Henan Normal University under contract No.11112, and the Joint Funds of the National Natural Science Foundation of China (U1404113).

\end{document}